\documentstyle[12pt,aaspp4,psfig]{article}
\def\be{\begin{equation}}
\def\ee{\end{equation}}
\def\etal{{\it et al. }}
\def\kms{km~s$^{-1}$~}

\def\Hnot{H$_\circ$}
\def\nrin{374~}
\def\nin+{584~}

\def\W50{W$_{50}$~}
\received{Someday 1999}
\accepted{Someday 1999}
\journalid{337}{Someday 1999}
\articleid{11}{14}

\slugcomment{Submitted {\it The Astronomical Journal}}

\begin{document}
\title{The Motions of Clusters of Galaxies and the Dipoles 
of the Peculiar Velocity Field}

\author {Riccardo Giovanelli, Martha P. Haynes}
\affil{Center for Radiophysics and Space Research
and National Astronomy and Ionosphere Center\altaffilmark{1},
Cornell University, Ithaca, NY 14853}

\author {John J. Salzer}
\affil{Astronomy Dept., Wesleyan University, Middletown, CT 06459}

\author {Gary Wegner}
\affil{Dept. of Physics and Astronomy, Dartmouth College, Hanover, 
NH 03755}

\author {Luiz N. da Costa}
\affil{European Southern Observatory, Karl--Schwarzschild--Str. 2, D--85748
Garching b. M\"unchen, Germany and Observatorio Nacional, Rio de Janeiro,
Brazil}

\author {Wolfram Freudling}
\affil{Space Telescope--European Coordinating Facility and European Southern 
Observatory, Karl--Schwarzschild--Str. 2, D--85748 Garching b. M\"unchen, 
Germany}

\altaffiltext{1}{The National Astronomy and Ionosphere Center is
operated by Cornell University under a cooperative agreement with the
National Science Foundation.}

\hsize 6.5 truein
\begin{abstract}

In preceding papers of this series, TF relations for galaxies in 24 
clusters with radial velocities between 1000 and 9200 \kms (SCI sample) were
obtained, a Tully--Fisher (TF) template relation was constructed and mean 
offsets of each cluster with respect to the template obtained. Here, an 
estimate of the line--of--sight peculiar velocities of the clusters and 
their associated errors are given. It is found that cluster peculiar 
velocities in the Cosmic Microwave Background reference frame
do not exceed 600 \kms and that their distribution has a line--of--sight 
dispersion of 300 \kms, suggesting a more quiescent cluster peculiar velocity 
field than previously reported. When measured in a reference frame in which 
the Local Group is at rest, the set of clusters at $cz > 3000$ \kms 
exhibits a dipole moment in agreement with that of the CMB, both in 
amplitude and apex direction. It is estimated that the bulk flow of a 
sphere of 6000 \kms radius in the CMB reference frame is between 140 and
320 \kms.

\end{abstract}

\keywords{galaxies: distances and redshifts  -- cosmology: 
observations; cosmic microwave background; distance scale }

\section{Introduction}

The Cosmic Microwave Background (CMB) radiation dipole moment is 
generally interpreted as a Doppler shift resulting from the motion 
of the Sun with respect to the comoving reference frame. The
vector associated with that motion has an amplitude of $368.7\pm2.5$ \kms,
and is directed toward $l=264.31^\circ\pm 0.16$, $b=+48.05^\circ\pm 0.09$
(Lineweaver \etal 1996). Allowing for solar motion with respect to the 
Local Standard of Rest, rotation of the Local Standard of Rest about
the galactic center and motion of the Galaxy with respect to the 
center of mass of the Local Group (LG) of galaxies (motions known
with increasing absolute uncertainty), the CMB dipole translates into
a velocity ${\bf V}_{cmb}$ of the LG with respect to the comoving 
reference frame, of amplitude $611\pm22$ \kms and directed towards 
$l=273^\circ\pm 3^\circ$, $b=27^\circ\pm 3^\circ$. Most of the 
uncertainty in the latter vector arises from that on the motion of
the Sun with respect to the LG, which we assume to have an amplitude
of 300 \kms and directed towards $l=90^\circ$, $b=0^\circ$ (de Vaucouleurs,
de Vaucouleurs \& Corwin 1976)\footnote{Several solutions of the solar
motion with respect to the LG exist. When expressed in terms of their Cartesian 
components --- directed respectively towards the Galactic Center, 
$(l,b)=(90^\circ,0^\circ)$ and  $b=90^\circ$ ---, the most frequently 
used solutions are respectively (i): $(0,300,0)$ (De Vaucouleurs \etal 1976), 
(ii): $(-79,295,-38)$ (Yahil, Sandage \& Tammann 1977) and (iii): $(-30,297,-27)$
(Lynden--Bell \& Lahav 1988), with units in \kms. They agree within the
accuracies with which each is determined, and we follow Lynden--Bell \& Lahav's
suggestion to adopt the easy--to--remember solution (i), for the sake of
standardization and simplicity. The vector differences of the Lineweaver 
\etal (1996) dipole and the above mentioned solutions yield estimates of 
the the LG motion with respect to the CMB. They are, respectively,
(i):   $(-24,-545,274)$ or 611 \kms towards $(l,b)=(273^\circ,+27^\circ)$;
(ii):  $(55,-540,312)$ or 626 \kms towards $(l,b)=(276^\circ,+30^\circ)$ and
(iii): $(6,-542,301)$ or 620 \kms towards $(l,b)=(271^\circ,+29^\circ)$.}.

In linear theory, the peculiar velocity induced on the LG by the
inhomogeneities present within a sphere of radius $R$ is
\be
{\bf V}_{pec,LG}(R) = {H_\circ \Omega_\circ^{0.6}\over 4\pi} 
\int \delta_{mass}({\bf r})
{{\bf r}\over r^3} W(r,R) d^3{\bf r}     
\ee
where $W(r,R)$ is a window function of width $R$, $H_\circ {\bf r}$ is
the distance in \kms, $\delta_{mass}$ is the mass overdensity at {\bf r}
and $\Omega_\circ$ is the cosmological density parameter. If the 
CMB dipole is the result of a Doppler shift, as we will assume in the 
remainder of this paper, then there must be identity between 
${\bf V}_{cmb}$ and ${\bf V}_{pec,LG}(R)$ as $R\rightarrow \infty$.
Direct measurements of the peculiar velocity field of galaxies and
clusters allow us such a comparison. They also allow an estimate of
the {\it convergence depth} of the local Universe. The integral in 
Equation (1) converges in the measure in which the average value
of $\delta_{mass}$ within a shell of radius $R$ approaches zero, as 
$R$ increases. In a universe which on large scales is homogeneous,
the convergence depth is approached at scales several times larger than
the correlation scalelength. In a fractal universe, the issue is more
complex (Pietronero \etal 1997; Guzzo 1998). We define the convergence
depth $d_c$ as the distance at which
first ${\bf V}_{pec,LG}(R)={\bf V}_{cmb}$, within the errors. Because
${\bf V}_{pec,LG}(R)$ may oscillate before settling on an asymptotic 
value, as suggested by the results of Hudson (1993), Strauss \etal (1992),
Scaramella \etal (1994), Tini--Brunozzi \etal (1995) and
Branchini \etal (1996), among others, the concept of convergence
depth is somewhat ambiguous. Nonetheless, since the radius sampled
by the clusters in the sample discussed in this paper is larger than 
the correlation length of the galaxian population as obtained from
redshift surveys, it is
legitimate to ask whether the peculiar velocity induced by the
large--scale distribution of matter they trace approaches 
${\bf V}_{cmb}$. In other words, is convergence reached within the
largest scale sampled by the cluster sample?

The first attempt to measure the large scale motion of the LG was
carried out by Rubin \etal (1976). For an all--sky sample of 96
Sc I galaxies enclosed in the redshift shell bound by 3500 and 6500
\kms, they measured the dipole of the quantity $HM = \log (cz)
-0.2 m$, where $z$ is the redshift, $m$ the galaxy's apparent magnitude 
and $c$ the speed of light. The derived dipole suggested a LG motion of
$454\pm 125$ \kms towards $(l_{Sc I},b_{Sc I})=(163^\circ,-11^\circ)$, 
significantly discrepant with that indicated by 
${\bf V}_{cmb}$, which was measured soon thereafter.

Virgocentric infall (as recently redetermined by Jerjen \& Tammann 1993),
contributes only a fraction of the motion of the LG. The amplitude of the
LG infall towards Virgo, which is directed about 45$^\circ$ away from
${\bf V}_{cmb}$, is on the order of 200 \kms. Shaya (1984) and Tammann
\& Sandage (1985) suggested that the Hydra--Centaurus supercluster, at a 
redshift of 3--4000 \kms, or supercluster structures obscured in the Zone 
of Avoidance (ZoA), played a more important role than Virgo in determining
the peculiar velocity of the LG. This suggestion found confirmation
in the analysis of Lynden--Bell \etal (1988), who proposed the
existence of a ``Great Attractor'' (GA) located at 4350 \kms and very
close to the galactic plane. Applications of the {\it Potent}
method yield density field reconstructions which, albeit grossly smoothed,
exhibit a broad density peak near $(l_{ga},b_{ga})=(320^\circ,0^\circ)$ 
at 4000 \kms (Dekel 1994; da Costa \etal 1996).
In 1989, Scaramella \etal pointed out the directional coincidence between
the GA and the Shapley Supercluster, a large concentration of clusters
near $cz \sim 14000$ \kms, hypothesizing that infall towards such distant 
structure is an important component of the local peculiar velocity field.
This result was echoed by the interpretations of Willick (1990) and
Mathewson \etal (1992) and disputed by Dressler \& Faber (1990). More
recently, Lauer \& Postman (1994) reported that the LG motion 
with respect to the reference frame defined by a sample of 119 clusters of 
galaxies extending to $cz \sim 15000$ \kms can be represented by a vector 
${\bf V}_{lp}$ of amplitude $561\pm284$ \kms, directed towards 
$(l,b)=(220^\circ,-28^\circ)\pm 27^\circ$.
The large discrepancy between ${\bf V}_{lp}$ and ${\bf V}_{cmb}$ was
interpreted as due to an overall bulk flow of the cluster reference
frame of $689\pm178$ \kms towards $(l,b)=(343^\circ,+53^\circ)$. The
dynamical implication of this result, which was confirmed by neither
Riess \etal (1995) nor Giovanelli \etal (1996), is that the LG motion
and that of the Lauer \& Postman cluster sample are caused {\it largely} 
by mass concentrations beyond 15,000 \kms, thus postulating a very large, 
or altogether absence of, local convergence depth.

This paper is part of a series based on spectroscopy and I band 
photometry of spiral galaxies, obtained with the purpose of improving 
the calibration of the Tully--Fisher relation (Tully and Fisher 
1977) and our understanding of the peculiar velocity field in the
local universe. Previous papers in the series are listed in 
Giovanelli \etal (1997a,b, hereafter Papers VI and VII). In Paper VI
we presented a set of galaxy TF parameters in cluster fields (SCI sample), 
and in Paper VII we obtained a template TF relation by combining the 
cluster data. The general motivations for those studies are given in the 
introduction of Paper VI.

The main goal of this paper is to investigate the large--scale 
deviations from Hubble flow, as traced by the clusters of galaxies
introduced in Papers VI and VII. Although the set includes only 24
clusters, their peculiar velocities are derivable with relatively
high accuracy. The analysis we carry out is simple and, in
comparison with that made with samples of field galaxies,
its results require much smaller corrections for observational
biases. Analogous studies have been carried out in the past
(e.g. Mould \etal 1991), but with data of inferior
accuracy (see Scodeggio 1997). Due to the sparse sampling provided
by clusters, the results of this study do not provide a detailed 
description of the peculiar velocity field. However, they can yield 
a good estimate of its dipole moment and allow a direct comparison 
with ${\bf V}_{cmb}$.

It is well known that the estimate of peculiar velocities via the TF
relation is independent on the assumed value of the Hubble constant 
\Hnot. The technique does however need careful calibration in order 
for the {\it velocity zero point} to be established. The template
relation obtained by us in Paper VII provides such calibration. In
that work, we also estimated the departure in magnitudes of each cluster 
TF relation from the template relation, after correction for a number of 
biases that arise in this type of analysis. If the calibration of the 
template relation is correct, the magnitude offset of each cluster can 
be combined with its systemic velocity to estimate a peculiar velocity. 
In Section 2, we thus estimate cluster motions. If the calibration of the 
TF template relation is incorrect, a geocentric component in the derived 
peculiar velocity field is introduced, simulating either 
global expansion or contraction. This would not alter the dipole 
signature of the peculiar velocity field but the individual velocities 
would be obviously incorrect. In Section 3 we discuss the accuracy of 
our template relation and thus estimate the degree to which our TF 
template relation can be globally assumed to define rest with respect 
to the comoving reference frame. In Section 4 we discuss the 
global motion of such large--scale structures as the Pisces--Perseus
supercluster (PP) and the Hydra--Centaurus supercluster (HC).
In Section 5 we inspect the incompleteness of our cluster sample  
and discuss its impact on
the derivation of dipole parameters. In Section 6, we derive the
cluster peculiar velocity distribution function. In Section 7 we
compute the dipole moment of the cluster peculiar velocity field
and investigate the amplitude of bulk flows in the
local universe. A summary of our results is presented in Section 8.

\section{Cluster Motions}                              

In Paper VI, we presented the TF parameters of the SCI sample of galaxies 
in 24 clusters. 
The clusters are well distributed over the sky and extend to a maximum 
$cz \sim 9200$ \kms. The relevant cluster parameters are given in Table 1
of Paper VI, while the galaxy data are listed in Table 2, ibidem. Cluster 
membership was assigned to each galaxy in the sample on the basis of 
criteria that were more stringent than generally adopted in previous TF 
work. The photometric observations were all carried out in the I band. 

\begin{deluxetable}{lrrrcccrcr}
\tablenum{1}
\tablecaption{Cluster Motions}
\tablehead{
\colhead{Cluster}                   & \colhead{$<\Delta m>$}         &  
\colhead{$<\Delta m>$}        & \colhead{$cz_{hel}$}                 & \colhead{$cz_{LG}$} &  
\colhead{$cz_{cmb}$}                 & \colhead{$V_{tf,cmb}$}  & \colhead{$V_{pec,cmb}$} &
\colhead{$V_{tf,cmb}$} & \colhead{$V_{pec,cmb}$}      
\nl
\colhead{} & \colhead{{\bf in}} & \colhead{{\bf in+}} & \colhead{} & \colhead{} & \colhead{} & \colhead{{\bf in}} & 
\colhead{{\bf in}} & \colhead{{\bf in+}} & \colhead{{\bf in+}} \nl
\colhead{(1)} & \colhead{(2)} & \colhead{(3)} & \colhead{(4)} & \colhead{(5)} &
\colhead{(6)} & \colhead{(7)} & \colhead{(8)} & \colhead{(9)} & \colhead{(10)}  
}
\startdata
N383      & $-0.027(126)$ & $+0.003(076)$ & $5161(032)$ & 5368 & 4865 & 4805 & $ +60(279)$ & 4871 & $  -6(170)$ \nl
N507      & $-0.112(111)$ & $-0.043(094)$ & $5091(099)$ & 5291 & 4808 & 4566 & $+242(233)$ & 4714 & $ +94(204)$ \nl
A262      & $+0.015(085)$ & $-0.033(063)$ & $4918(080)$ & 5105 & 4664 & 4696 & $-32(+184)$ & 4594 & $ +70(133)$ \nl
A400      & $+0.077(111)$ & $+0.039(070)$ & $7142(075)$ & 7178 & 6934 & 7184 & $-250(367)$ & 7060 & $-126(227)$ \nl
Eridanus  & $+0.433(116)$ & $+0.393(087)$ & $1665(030)$ & 1567 & 1534 & 1872 & $-338(100)$ & 1838 & $-304(074)$ \nl
Fornax    & $+0.053(098)$ & $+0.173(090)$ & $1415(045)$ & 1266 & 1321 & 1354 & $ -33(061)$ & 1430 & $-109(060)$ \nl
Cancer    & $-0.113(092)$ & $-0.027(077)$ & $4705(080)$ & 4604 & 4939 & 4689 & $+250(198)$ & 4878 & $ +61(172)$ \nl
Antlia    & $-0.128(107)$ & $-0.133(080)$ & $2800(100)$ & 2517 & 3120 & 2941 & $+179(145)$ & 2935 & $+185(109)$ \nl
Hydra     & $+0.236(080)$ & $+0.164(070)$ & $3733(050)$ & 3465 & 4075 & 4542 & $-467(167)$ & 4395 & $-320(142)$ \nl
N3557     & $-0.052(136)$ & $-0.134(108)$ & $3000(070)$ & 2726 & 3318 & 3239 & $ +79(203)$ & 3119 & $+199(155)$ \nl
A1367     & $-0.014(065)$ & $-0.020(062)$ & $6408(088)$ & 6336 & 6735 & 6692 & $ +43(200)$ & 6673 & $ +62(191)$ \nl
Ursa Major& $+0.687(080)$ & $+0.710(080)$ & $ 896(040)$ &  965 & 1101 & 1511 & $-410(056)$ & 1526 & $-425(056)$ \nl
Cen30     & $-0.138(098)$ & $-0.213(070)$ & $3041(150)$ & 2805 & 3322 & 3117 & $+205(140)$ & 3012 & $+310(098)$ \nl
A1656     & $-0.052(065)$ & $-0.065(058)$ & $6917(068)$ & 6926 & 7185 & 7015 & $+170(210)$ & 6973 & $+212(186)$ \nl
ESO508    & $-0.382(185)$ & $-0.302(100)$ & $2900(100)$ & 2720 & 3210 & 2693 & $+517(230)$ & 2793 & $+417(128)$ \nl
A3574     & $+0.073(165)$ & $+0.012(078)$ & $4548(030)$ & 4374 & 4817 & 4982 & $-165(379)$ & 4843 & $ -26(174)$ \nl
A2197\dag & $+0.066(160)$ & $+0.048(090)$ & $9138(100)$ & 9334 & 9162 & 9444 & $-282(693)$ & 9285 & $-206(384)$ \nl
A2199\dag & $+0.056(160)$ & $+0.048(090)$ & $8970(098)$ & 9163 & 8996 & 9231 & $-235(681)$ & 9285 & $-206(384)$ \nl
Pavo II   & $-0.118(175)$ & $-0.068(082)$ & $4470(070)$ & 4342 & 4444 & 4209 & $+235(339)$ & 4307 & $+137(163)$ \nl
Pavo      & $-0.108(225)$ & $-0.043(120)$ & $4100(100)$ & 3952 & 4055 & 3858 & $+197(400)$ & 3975 & $ +80(219)$ \nl
MDL59     & $+0.487(180)$ & $+0.427(092)$ & $2590(075)$ & 2636 & 2317 & 2900 & $-583(240)$ & 2820 & $-503(120)$ \nl
Pegasus   & $+0.032(126)$ & $+0.112(105)$ & $3888(080)$ & 4087 & 3519 & 3571 & $ -52(207)$ & 3705 & $-186(180)$ \nl
A2634\ddag& $+0.020(070)$ & $+0.033(065)$ & $9240(079)$ & 9484 & 8895 & 8977 & $ -82(289)$ & 9031 & $-136(270)$ \nl
A2666     & $+0.043(127)$ &               & $8118(081)$ & 8357 & 7776 & 7931 & $-156(459)$ &      &             \nl
\enddata
\tablenotetext{\dag} {Parms. for {\bf in+} samples include A2197, A2199 and peripheral objects.}
\tablenotetext{\ddag}{Uses expanded sample of Scodeggio, Giovanelli and Haynes (1997).}
\end{deluxetable}

Galaxies associated with each cluster are separated into two classes: 
(i) the {\bf in} sample includes galaxies that are very likely cluster 
members, on the basis of their sky and redshift coordinates; (ii) a class 
of ``peripheral'' galaxies is characterized by redshift quite close to the 
systemic one of the cluster, but sufficiently removed spatially from the 
cluster center so that a membership assignment cannot be reliably made. The 
combination of {\bf in} and peripheral objects for a given cluster is 
referred to as the {\bf in+} sample. The SCI {\bf in} sample of the 24 
clusters includes \nrin galaxies, of which 360 are deemed good candidates
for TF work, while the SCI {\bf in+} sample includes \nin+ objects, of 
which 555 are used for TF work. The remaining 198 galaxies for which TF 
parameters were presented in Paper VI are either foreground or background 
objects, or members of inadequately sampled groups/clusters.

In Paper VII, the cluster data were combined to obtain a template TF
relation. This was done separately for the {\bf in} and for the {\bf in+} 
objects. To that end, the subset of 14 clusters with $cz > 4000$ \kms was 
assumed to yield a null average peculiar velocity. This is equivalent to
setting equal to zero the amplitude of any {\it geocentric} global deviation
from Hubble flow for the spherical shell between 4000 and 9200 \kms, and 
translates into the definition of a zero point of the template TF relation. 
Magnitude offsets $\Delta m$ of each cluster with respect to that template 
can be converted into peculiar velocities via the relation
\be
V_{pec,cmb} = cz_{cmb} (1 - 10^{0.2\Delta m})   
\ee
where $cz_{cmb}$ is the cluster systemic velocity with respect to the
CMB reference frame and $V_{pec,cmb}$ is the cluster peculiar velocity,
in the same reference frame.

In Table 1, we list the mean magnitude offsets $\Delta m$, computed 
separately for each ({\bf in}) and each ({\bf in+}) cluster sample,
respectively in columns 2 and 3. Those offsets are derived as 
\be
\Delta m = (1/4) \sum_{i=1}^4 \Delta y_i,   
\ee
where the $\Delta y_i$ are the magnitude offsets listed in columns 4--7
of Table 3 of Paper VII; the four $\Delta y_i$'s were estimated for
two different slopes of the faint end of the galaxy luminosity function
(each leading to different incompleteness bias corrections for the various
cluster samples) and for both a linear and a quadratic fit to the TF template 
relation. The four solutions differ slightly, by amounts generally
smaller than the typical uncertainty. An equal--weight average of the four 
offsets is adopted here. The uncertainty on $\Delta m$, $\epsilon_\mu$, is 
listed in parenthesized form in Table 1: i.e. $-0.027(126)$ is equivalent
to $-0.027\pm 0.126$. $\epsilon_\mu$ is an error  on the TF distance 
modulus and the cumulative result of the uncertainties arising from:
 
\noindent (i) the number of galaxies and the quality of their TF parameters 
in each cluster sample; 

\noindent (ii) the TF template relation parameters, both
those deriving from the formal fits and those associated with systematic
effects, as gauged in Paper VII, Section 6; 

\noindent (iii) the cluster incompleteness bias correction (see Paper VII, 
Section 6.3); 

\noindent (iv) the measured systemic velocity of each cluster; 

\noindent (v) the quality of the kinematical zero point, obtained by assuming 
that the mean peculiar velocity of clusters at $cz_{cmb}>4000$ \kms is zero, 
as described in Section 6.2 of Paper VII.  

\noindent The values $\epsilon_\mu$ listed in Table 1 are conservative, erring 
more likely toward an overestimate of the uncertainties, except for an important
proviso. The uncertainty arising from the quality of the kinematical zero point, 
mentioned in (v) above, is derived from the amplitude of the cluster peculiar
velocity distribution function and on the number of objects used, on the
assumption that the mean peculiar velocity of distant clusters is zero, i.e.
that there is no large--scale geocentric signature in the peculiar velocity
field. If however the LG were to be located near the center of a large--scale,
isotropic void or positive density enhancement, an acceleration or
delay of the Hubble expansion would take place, and the peculiar velocity
field would have a geocentric signature. We discuss this possibility in Section
3. For the moment, we assume that such a signature is of negligible amplitude. 

Other contents of Table 1 include:
the systemic velocity of the cluster in the solar, Local Group and CMB
reference frame (columns 4, 5 and 6); the peculiar velocities measured
in the CMB reference frame for each cluster, respectively for the {\bf in}
(column 8) and the {\bf in+} (column 10) samples;  the
TF distance of the cluster, expressed in \kms, i.e.
\be
cz_{tf,cmb} = cz_{cmb} - V_{pec,cmb}    
\ee
is tabulated for the {\bf in} sample in column 7 and the {\bf in+} sample in 
column 9. Peculiar velocities and TF distances in Table 1 are not corrected
for the Malmquist bias, which is discussed in Section 6. Errors on the
systemic and peculiar velocities of the clusters are given in parenthesized
form in columns 4, 8 and 10. Errors on the peculiar velocities, inferred 
from $\epsilon_\mu$,
are actually slightly asymmetric about the mean value adopted; for example,
$\Delta m = +0.015\pm 0.085$ for the {\bf in} sample of A262 translates into
$V_{pec,cmb} = -32^{+180}_{-188}$. In Table 1, we list the uncertainty on
$V_{pec,cmb}$ as the mean of the upper and lower side errors.

Note that in the case of the clusters A2197 and A2199, a single {\bf in+}
sample is adopted, which includes both clusters, very close in redshift to
each other, and their peripheries. For A2666, no {\bf in+} sample is defined, 
due to the confusing nearness in projection of A2634. 



For A2634, rather than the samples described in Paper VII
we use those presented by Scodeggio, Giovanelli \& Haynes (1997),
which include additional objects, unavailable for the Paper VII study.
The expansion of the A2634 sample size by approximately one third 
leads to a revision of the average magnitude offset, with respect to
the TF template relation, of -0.035 mag, in comparison to the Paper VII
result, a change amounting to about $(1/2)\epsilon_\mu$.

A few interesting details of Table 1 are worth underscoring. First,
no values of the cluster line--of--sight peculiar velocity are measured, 
in excess of $\pm 600$ \kms, in the CMB reference frame. Second, the
largest velocities are measured for nearby groups, structures within
$cz\sim 3000$ \kms from the Local Group. This overall relatively quiescent
picture of the peculiar velocity field is in marked contrast of previous
cluster peculiar velocity measurements. The comparison between the
peculiar velocities obtained from spiral and elliptical samples for a
single cluster, as shown in Mould \etal (1991) was until recently
discouragingly poor, suggesting that the large amplitude of the estimated
peculiar velocities resulted from systematic errors unaccounted for by
the reported scatter in the TF or Fundamental Plane techniques. The 
situation has improved significantly, and recent
comparisons exhibit very noticeably reduced scatter, a higher degree of 
correlation and much reduced amplitude in the peculiar velocities inferred, 
as shown by Scodeggio, Giovanelli \& Haynes (1998).

\section {Does the TF Template Relation Represent 
a Rest Reference Frame?}                               

The TF relation is a linear function relating the logarithm of a
galaxy's velocity width and its absolute magnitude: 
\be
M = a + b (\log W-2.5).   
\ee
In paper VII, we estimated the total uncertainty on $a$ to be on the order 
of 0.05 mag, arising mostly from the limits on our ability to characterize 
the ensemble of clusters with $cz_{tf}>4000$ \kms as a good approximation 
to a comoving reference frame. The uncertainty on the slope $b$ is on the 
order of 2\%. 

An error on the zero point $a$ of the TF template relation translates 
into the spurious generation of a geocentric peculiar velocity field, of 
amplitude which increases linearly with distance. In other words, it 
simulates an isotropic distortion $\delta H$ of the Hubble expansion. An 
error of 0.05 mag in $a$ simulates a slowdown or speeding up of the Hubble 
expansion by 2.3\%. 

An error on the slope $b$ also translates in an isotropic, spurious 
distortion of the Hubble flow. The distortion does however affect near and 
distant sources in different ways. If $a$ is correct and $b$ is too steep, 
for example, galaxies of large width will tend to have positive magnitude 
offsets $\Delta m$ with respect to (4), i.e. negative peculiar velocities,
while the opposite will be true for galaxies of velocity width $\log W < 2.5$.
Most TF samples span a large range of distances. At some distance $d_e$ within 
that range, galaxies of large width may be as likely to be members of the sample 
as galaxies of small width. The error in $b$ would then produce a spurious 
acceleration of the Hubble flow at $d<d_e$ and a spurious deceleration at $d>d_e$. 

These effects are easily spotted in all--sky surveys, while 
in surveys that concentrate on selected parts of the sky, such geocentric
effects are more difficult to identify and can easily be misinterpreted
as the signature of bulk flows. There is however an insidious possibility
that can wreak havoc even when full--sky coverage is available. If, as
briefly mentioned in the preceding Section, the LG were to be located
near the center of a roughly spherical, large--scale density fluctuation,
geocentric distortions of the Hubble flow would be real, rather than the
result of poor parametrization of the distance determination technique.
Recently, Zehavi \etal (1998) have suggested precisely that possibility:
that the volume within $cz\sim 7000$ \kms is subject to a Hubble acceleration
of $(6.6\pm2.2)$\%, resulting from a local underdensity of 20\%, surrounded
by an overdense shell. The possibility of a large--scale geocentric peculiar 
velocity field was excluded by the way our template TF relation was defined.
However, the Zehavi \etal result can in principle be tested with our data: a 
``Hubble bubble'' can be distinguished from a distance calibration error by 
detecting the ``edge'' of the perturbed region. If present, the claimed effect 
would produce a differential TF offset of 0.14 mag between nearby clusters and 
those more distant than 7000 \kms.

In figure 1 we carry out such a test. Using the data in Table 1, we plot 
$\delta H/H = V_{pec,cmb}/cz_{tf,cmb}$ against $hd=cz_{tf,cmb}/100$
(with the implicit usual parametrization $H_\circ = 100 h$),
separately for the peculiar velocities computed for the {\bf in} and for
the {\bf in+} samples (panels $a$ and $b$ respectively). Inset in the figures 
are the average values of $\delta H/H$ for three intervals of $hd$:
30 to 60, 35 to 60 and 60 to 95 Mpc. At small distances, the peculiar
velocity field is quite unstable, dominated by the large velocities of
nearby groups, which are comparable to those of the LG (611 \kms) and constitute
a significant fraction of the Hubble flow. At distances larger than $35h^{-1}$ 
Mpc, the monopole of the cluster velocity field does not exhibit significant 
change of value, the Hubble flow at $hd>60$ Mpc appearing to differ
from that between 30 and 60 Mpc by less than 3\%, rather than the 6.6\%
reported by Zehavi from their SNe Ia sample. The results shown in Figure 1
should however be considered inconclusive, as the redshift range of our 
sample is about three times smaller than that of the SNe Ia, and our cluster
sample barely straddles the edge of the purported change of regime on the 
Hubble flow. The small number of clusters farther than $70h^{-1}$ Mpc
allows for cosmic variance to mask the effect of a change in the Hubble
flow. While our data do not corroborate the claim of Zehavi \etal, neither 
do they refute it. A more thorough check of the Zehavi \etal
hypothesis will be possible soon, as the Dale \etal (1997a,b) survey of
TF distances of clusters to $cz\simeq 20000$ \kms is completed.

If we allow the kind of ``Hubble bubble'' effect claimed by Zehavi \etal,
our estimate of the uncertainty on $a$ would increase somewhat. Because
the clusters in our sample straddle the edge of the bubble, the impact
of the geocentric flow on $a$ is not very large. Given the cluster
distance distribution, we estimate that the net shift in the TF template
magnitude offset, between a ``no geocentric flow'' solution and a
``Hubble bubble'' solution, to amount to less than 0.03 mag. The 
combination of this contribution with the already quoted uncertainty
of 0.05 mag, would raise the error on $a$ to 0.06 mag. Zehavi \etal
have cogently argued that a region of $70^{-1}$ Mpc radius could be
underdense by 20\% or so, which is the amount necessary to produce
the suggested local acceleration of the Hubble flow, without stretching
too hardly the range of plausibility of the cosmological power spectra.
One would be left, of course, with the nagging coincidence of the central 
location of the LG in the void. But non--conspiratorial coincidences do occur.

If the LG were to be near the center of a ``Hubble bubble'', dipole
solutions would not be affected, whether the bubble effect were to be
maintained in the data or removed by the exclusion of geocentric
solutions. The peculiar velocities measured for individual clusters, on 
the other hand would differ in the two scenarios: the exclusion of 
geocentric solutions would reduce the peculiar velocity estimates, 
producing a somewhat more quiescent picture of the kinematic fluctuations.
In the remainder of this paper, we will carry out calculations using
the peculiar velocities as listed in Table 1. When applicable, the
effect of a ''Hubble bubble'' on our results will be estimated and
discussed.

\section {The Global Motion of Supercluster Structures}   

Our cluster sample yields interesting information on the global motion
of three large--scale structures: the Perseus--Pisces, Coma and 
Hydra--Centaurus superclusters.
The two main clusters in the Coma region, A1367 and A1656, both have small
peculiar velocities which average to about $+150$ \kms; within one--sigma,
the global motion of the supercluster with respect to the CMB reference 
frame is nil. 

The two Pisces groups (N383 and N507) and A262 in the Perseus--Pisces 
supercluster also exhibit a similar pattern: none has a large amplitude 
$V_{pec}$ and the average of their motions is about 75 \kms; within 
one--sigma, the global motion of the supercluster is again nil. It has been 
claimed (Willick 1990) that Perseus--Pisces as a whole has a large 
negative velocity, on the order of -400 \kms in the CMB reference frame; 
this is of comparable amplitude to the velocity of infall of the LG towards 
the GA region, suggesting that both the LG and Perseus--Pisces are `travel
companions' in their infall towards a structure much more distant than 
Hydra--Centaurus (Scaramella \etal 1989). Our results indicate that the 
denser regions in the Perseus--Pisces supercluster do not have a global 
motion as large as 400 \kms with respect to the CMB, at the better than 
99\% confidence level, and therefore that the region between the LG and 
the Perseus--Pisces supercluster is affected by a relatively
steep peculiar velocity gradient.
  
The GA region, intended loosely as the conglomerate of groups and clusters 
which includes Antlia, N3557, Cen30, Hydra, ESO 508 and A3574 among objects 
in our sample, presents a more complex picture. It lacks a clear, large
amplitude central structure and it stretches in $cz_{tf}$ between 2500
and 5000 \kms. The cluster A3627, located at $(l,b)=325^\circ,-7^\circ)$
at $cz\simeq 4300$ \kms and studied by Kraan--Korteweg \etal (1996),
is at too low a galactic latitude to permit the accurate photometry
required by TF work. The foreground structures, Antlia, ESO508 and Cen30,
exhibit significant outflow (positive) velocities, while the two structures 
in the background, A3574 and more significantly Hydra, exhibit backflow 
(negative) velocities. The large amplitude of the velocities of several of 
the clusters, between 200 and 450 \kms, are significant at the 2--3--sigma 
level. While the evidence of the motion of Hydra and A3574 alone may be 
statistically too sparse to allow a claim of backflow in the GA region, it 
tends to corroborate rather than refute the early claim of Dressler \& 
Faber (1990). The observation of a negative velocity for Hydra suggests 
that the early suggestion of Shaya (1984) and Tammann \& Sandage (1985), 
that centered the successively named GA no farther than Hydra--Centaurus, 
may be correct. Overall, the scenario that emerges from the data in Table 
1 is one where the large--scale structures in the local universe exhibit 
very little global deviation from smooth Hubble flow.

\section {Completeness of the Cluster Sample}  

Our cluster sample is not a fair sample of the local universe. At low
redshifts it includes several groups of low enough richness so that
they do not meet criteria for inclusion in the Abell/ACO catalogs. Beyond
$cz\sim 6000$ \kms, the majority of Abell/ACO clusters are not included 
in our sample. Figure 2 
%
%
shows a redshift histogram of our cluster sample vis--a--vis with that
of the members of the Abell catalog. Between $cz \simeq 2000$ \kms and 
$cz\sim 9000$ \kms, the selection function $s(cz)$, i.e. 
the probability of a given Abell/ACO cluster to be included in our sample,
drops roughly in proportion to $e^{-(cz/2300)}$. Only one in ten Abell
clusters with $cz$ between 7000 and 9500 \kms enters our sample.
The steepness of the selection function has an impact on the
kinematical inferences discussed in this paper. For example, the
estimate of bulk flows in the peculiar velocity field, and especially 
its comparison with the results of cosmological simulations, generally 
refers to the global motion of the matter within a region bounded by a 
top hat or a Gaussian filter, more frequently the former. In an observed
sample such as ours, it would then be necessary to assign higher weight
to distant clusters, roughly by a factor proportional to the inverse
of the selection function, in order to obtain estimates not inordinately
affected by the characteristics of the very local peculiar velocity
field. This form of weighting increases the uncertainty of derived 
dipoles, because the peculiar velocity of clusters is known with an 
accuracy $\epsilon_v$ which is roughly of the order of 4\% of their 
$cz_{tf}$. In an volume--weighted measure of the peculiar velocity 
dipole or bulk flow, 
each object should be weighted according to $(s \epsilon_v^2)^{-1}$,
which in our case turns out to be a roughly constant value between 2000 
and 9000 \kms, the redshift stretch of our cluster sample. In other
words, the computation of a bulk flow or a dipole using {\it equal
weights} for distant and nearby clusters approximates filtering the
peculiar velocity field by a top hat of radius equal to the redshift
range of our data.

\section {The Cluster Peculiar Velocity Distribution Function} 

It is of interest to know the distribution function of peculiar velocities
of clusters. Such function can provide an indication of the variance in the
peculiar velocity field, and it can be directly compared with numerical
simulations obtained within the framework of different cosmological models,
thus providing an observational test for their adequacy. Before we proceed
to an estimate of the distribution function, we touch the problem of
Malmquist bias on peculiar velocity measurements.

Statistical underestimates of TF distances arise due to the Malmquist bias, 
a well--known effect which results from the fact that distance measurements
are uncertain and that within a given solid angle the number of possible 
targets with distance between $r$ and $r+dr$ usually increases with $r$. 
Thus, for a set of targets of estimated distance modulus $\mu_e \pm \epsilon_\mu$
the most probable distance is not $r_e = 10^{0.2(\mu_e-c)}$ (where $c$ is
the usual scaling term that depends on the adopted units of distance), but
a value $r>r_e$, because the distribution of targets is such that there is
a larger number of them between $\mu_e$ and $\mu_e + \epsilon_\mu$ than between
$\mu_e$ and $\mu_e - \epsilon_\mu$. When the assumption is made that the targets
are distributed in space in Poisson form, the effect is referred to as the
``homogeneous Malmquist bias'' (HMB). In that case,
\be
r = r_e e^{3.5\Delta^2} \qquad\qquad {\rm where}\qquad\qquad    
\Delta=10^{0.2\epsilon_\mu} - 1
\ee
$\Delta$ is the relative distance error. When the targets are individual 
galaxies, the TF relation yields distance moduli with an uncertainty on the 
order of $\epsilon_\mu \simeq 0.3$
mag, which translates in a HMB correction $r/r_e \simeq 1.08$.
For clusters, the uncertainty $\epsilon_\mu$ is significantly reduced, and $r/r_e$
is closer to 1.01.

The HMB modifies the TF distances $cz_{tf}$ in columns 7 and 9 of Table 1
to $cz_{tf}|_{mb}$, as shown in Equation (6), where the errors $\epsilon_\mu$ 
are those listed in columns 2 and 3 of Table 1. The peculiar velocities are 
thus modified from the values listed in Table 1 to 
$V_{pec}|_{mb}=cz - cz_{tf}|_{mb}$. The corrections 
are generally small, except for the most distant clusters or those with large 
$\epsilon_\mu$. We use such modified values in the estimate of the distribution 
function of peculiar velocities of clusters, but for simplicity we will
forego carrying the clumsy ${mb}$ subscript. Since the distribution of clusters 
in space is not Poissonian, the bias correction should in principle take
into account the clustering characteristics of the distribution. However,
since the corrections are small, the cluster population is quite sparse
and an inhomogeneous correction is difficult to estimate and thus highly
uncertain, the correction for 
a simple HMB is deemed sufficient for our purposes. In fact, both in the
computation of the peculiar velocity distribution function, presented in
the remainder of this section, and in that of dipoles, as described in the
next section, the Malmquist bias correction has very little impact on the
final results. 

We have only access to one component of a cluster's peculiar velocity, that
along the line of sight. In addition, as indicated in Table 1, that value is
generally known with a significant amount of uncertainty. In Figure 3, we 
graphically present the values and uncertainties of the peculiar velocities 
of the 24 clusters in our sample. The cluster peculiar velocities (for the 
{\bf in+} samples) are represented by Gaussian functions of equal area 
$A_i\exp [-(V_{pec,1d} - V_{pec,i})^2/2\epsilon_{v,i}^2]$ (where the 
amplitudes $A_i\propto \epsilon_{v,i}^{-1}$), centered at the value of 
$V_{pec,i}$ for the $i$--th cluster, and with dispersion equal to the 
error on $V_{pec,i}$, $\epsilon_{v,i}$. The sum of those yields the observed 
distribution function $f_{obs}(V_{pec,1d})$ of the line--of--sight peculiar
velocity values, measured in the CMB reference frame, as broadened by 
observational errors (heavy line in Figure 3, arbitrarily rescaled). The 
distribution is slightly asymmetric, due to the large velocities of nearby 
groups, such as Ursa Major and Eridanus. Note that $V_{pec}=0$ is defined by 
setting the monopole of clusters farther than 4000 \kms to zero, as 
discussed in Section 3.
A Gaussian fit with zero mean to $f_{obs}(V_{pec,1d})$ yields a dispersion
$\sigma_{1d,obs} = 325\pm54$ \kms. The uncertainty can be estimated from the 
nominal errors of the fit or, alternatively, by Monte Carlo simulations of 
synthetic data sets. The simulations are obtained by producing data sets 
where the peculiar velocity of the $i$--th cluster is a random deviate of 
$A_i\exp [-(V_{pec,1d} - V_{pec,i})^2/2\epsilon_{v,i}^2]$. Each simulation
of a cluster set is fit and the scatter among fitted values of $\sigma_{1d,obs}$
yields the uncertainty. The two estimates of error agree. Since the observed
distribution function is broadened by errors, $\sigma_{1d,obs}$ overestimates
the true dispersion. The broadening produced by the varied combination of
error functions is easily calculated by Monte Carlo simulations, yielding
a dispersion corrected for error broadening  $\sigma_{1d} = 270\pm54$ \kms.
These figures apply to the {\bf in+} samples. Repeating the same
exercise for the peculiar velocities obtained from the {\bf in} samples,
we obtain $\sigma_{1d,obs} = 375$ \kms and $\sigma_{1d} = 277\pm63$ \kms,
respectively.

What would be the effect of a Zehavi \etal (1998) Hubble bubble on our
estimate of $\sigma_{1d}$ ? We gauge it by allowing for an acceleration 
of the Hubble flow by 6\% within $cz\sim 7000$ \kms; we then estimate the 
shift in the kinematical zero point resulting from this assumption, with 
respect to that obtained if nil net flow for the clusters between 4000 
and 9500 \kms is assumed (which led to the peculiar velocities in Table 1). 
We correct the peculiar velocities by the zero point shift and repeat the 
calculations described in the previous paragraph. The inclusion of a Hubble
bubble broadens $\sigma_{1d}$ by 45 \kms. 

While the exclusion of large--scale geocentric flows may bias our estimate
of $\sigma_{1d}$ low, the presence in our sample of poor groups at low 
redshift (Ursa Major, Eridanus, MDL59) may have the effect of biasing the
result on the high side. Nearby groups have among the largest observed
peculiar velocities, which may be representative of locations in the 
peripheral parts of superclusters, regions characterized by high, 
large--scale density gradients. Rich clusters tend to reside in the denser 
parts of superclusters, near the bottom of gravitational potential wells, 
and the variance in their motions may be more subdued. 

Thus, allowing for uncertainties on systematic biases associated with 
a possible geocentric deviation from Hubble flow of a few percent and for 
the presence of 
small foreground groups in our sample, it appears that the cosmic variance 
in the 1--d peculiar velocity of clusters in the local universe can be well 
approximated by a r.m.s. value of the order of 
\be
\sigma_{1d} = 300\pm80\qquad{\rm km s}^{-1}  
\ee
This number is in good agreement with previous estimates by Bahcall \& Oh 
(1996) and by Watkins (1997), based on our data, and is significantly
lower than values obtained from previous measurements of cluster velocities
(e.g. Mould \etal 1991), which were affected by much larger errors than those
associated with this set. Bahcall and Oh (1996) and 
Borgani \etal (1997), compared the cluster peculiar distribution function 
obtained from these data with numerical simulations in the framework of 
different cosmological models, finding support in these data for models with 
relatively low values of $\Omega$.

\section {Dipoles of the Cluster Peculiar Velocity Field} 

\subsection {Procedures}  

Although few in number, the distance moduli of the clusters in our sample 
are determined with a high degree of accuracy, on the order of 0.08 magnitudes, 
or 4\% of $cz_{tf}$. As a result, a good estimate is possible of the low
order spherical harmonics of their peculiar velocity field, namely the dipole. 
We only measure the line of sight component of the peculiar velocity, thus
the problem reduces to computing the dipole of a scalar field. Since we are 
interested in the comparison with ${\bf V}_{cmb}$ and ${\bf V}_{lp}$, the 
apparent motions of the LG with respect to, respectively, the CMB and the 
Lauer \& Postman cluster reference frames, we shall estimate the dipole of 
the {\it reflex} motion of the LG with respect to our cluster set. 
If $-V_i$ is the peculiar velocity of the $i$--th cluster {\it in the LG
reference frame}, 
and $\epsilon_i$ is the uncertainty on that quantity, we solve for the
vector ${\bf V}_d$ of the dipole moment by minimizing the merit function
\be 
\chi^2 = \sum_i {1\over s_i}
\Big({V_i - {\bf V}_d \cdot {\bf \hat r}_i\over \epsilon_i}\Big)^2  
\ee
where $\hat {\bf r}_i$ is the unit vector in the direction of the $i$--th 
cluster and $s_i$ is the selection function at its distance.
The errors of a dipole solution based on a small number of samples $N_c$ 
can be capriciously distributed. We thus obtain an error estimate of our
results by producing a large number $N_{sets}$ of synthetic data sets, and 
monitoring the scatter among the resulting dipole solutions. Each synthetic
cluster set is obtained as follows:
the locations of the clusters, $[{\bf \hat r}_i]$, are maintained as
observed but the peculiar velocity of each cluster is extracted as a 
random deviate from a Gaussian of center $-V_i$ and dispersion $\epsilon_i$.
In our simulations, $N_{sets}=1000$. The error ellipsoid of the components
of the dipole solution of the observed cluster set is estimated from the 
scatter among the $N_{sets}$ dipole solutions for the synthetic cluster 
sets.

As discussed in Section 5, our cluster set is not a fair sample. Nearby
clusters are more likely to be part of the set than distant ones. An
estimate of the peculiar velocity dipole moment that does not take into 
consideration the selection function of the sample will thus heavily
weigh the nearby clusters over the more distant ones. As a function of
distance, the shape of the selection function $s$ and the lognormal character 
of the peculiar velocity errors do however combine in such a way that if 
each cluster's contribution to the dipole is weighed inversely proportional 
to the sample selection function, as in Equation (8), the product 
$s_i \epsilon_i^2$ is roughly constant. An alternative technique to 
correcting for the fading of the sample at higher redshifts is that of 
assigning a weight to each cluster which is proportional to $r^3_n$, where 
$r_n$ is the distance to the $n$--th nearest neighbor in the sample and $n$ 
is usually a number selected between 3 and 9. For a small sample such as ours,
this form of volume--weighting introduces a substantial measure of erratic
behavior, and for the purpose of approximating top hat volume--weighting, 
we adopt the simpler approach of using unit weights $s_i \epsilon_i^2$.

Below, we present dipole solutions using $(s_i) \equiv 1$, which give
large weight to nearby clusters (columns labeled `Case $a$' in Table 2), and
$(s_i \epsilon_i^2) \equiv 1$, which is equivalent to weighing each 
cluster in proportion inverse to the selection function discussed in
Section 5 (columns labeled `Case $b$' in Table 2). The second approach 
increases the effective depth of the solution, at the cost of increased 
noise. 

\begin{deluxetable}{lrrrrr}
\tablenum{2}
\tablecaption{Cluster Dipole Solutions}
\tablehead{
\colhead{}    & \colhead{}    &
\colhead{Case $a$:} & \colhead{$s_i\equiv 1$}            & 
\colhead{Case $b$:} & \colhead{$s_i\epsilon_i^2\equiv 1$} \nl
\tablevspace{5pt}
\colhead{Set} & \colhead{N$_c$} & \colhead{ V}  & \colhead{ $(l,b)$}       
                                      & \colhead{ V}  & \colhead{ $(l,b)$}  \nl
\tablevspace{5pt}
\colhead{} & \colhead{} & \colhead{km s$^{-1}$} & \colhead{$^\circ$} & \colhead{km s$^{-1}$} & \colhead{$^\circ$}  
}
\startdata
1. All  				        & 24 & 759$\pm$083 & (229,+31)$\pm$11 & 450$\pm$141 & (266,+31)$\pm$26 \nl
2. All ZOA $\bar V_{pec}=0$ 		        & 30 & 449$\pm$121 & (234,+43)$\pm$28 & 364$\pm$148 & (272,+38)$\pm$31 \nl
3. All ZOA $V_{pec}$ from SF 			& 30 & 609$\pm$100 & (240,+36)$\pm$18 & 472$\pm$118 & (266,+26)$\pm$22 \nl
\\
4. $cz_{tf}>3000$ 			        & 18 & 611$\pm$129 & (263,+18)$\pm$18 & 496$\pm$196 & (270,+37)$\pm$37 \nl
5. $cz_{tf}>3000$ ZOA $\bar V_{pec}=0$  	& 23 & 447$\pm$141 & (277,+48)$\pm$35 & 433$\pm$173 & (278,+51)$\pm$45 \nl
6. $cz_{tf}>3000$ ZOA $V_{pec}$(SFI)		& 23 & 565$\pm$103 & (268,+22)$\pm$17 & 484$\pm$158 & (271,+38)$\pm$30 \nl
\\
7. $cz_{tf}<6000$ 			        & 17 & 794$\pm$070 & (231,+31)$\pm$11 & 543$\pm$090 & (263,+23)$\pm$16 \nl
8. $cz_{tf}<6000$ ZOA $\bar V_{pec}=0$  	& 21 & 469$\pm$121 & (239,+43)$\pm$25 & 433$\pm$161 & (275,+34)$\pm$27 \nl
9. $cz_{tf}<6000$ ZOA $V_{pec}$(SFI)		& 21 & 663$\pm$083 & (239,+36)$\pm$17 & 534$\pm$090 & (268,+19)$\pm$16 \nl
\enddata
\end{deluxetable}

Equation (8) is solved for the three Cartesian components of ${\bf V}_d$,
directed respectively towards the Galactic Center, $(l,b)=(90^\circ,0^\circ)$
and the $b=90^\circ$. The amplitudes of the dipoles listed in Table 2 are 
corrected for the ``error bias'', i.e. 
$|{\bf V}_d|^2 = V_{dx}^2+V_{dy}^2+V_{dz}^2-e_x^2-e_y^2-e_z^2$, where 
$e_x$, $e_y$, $e_z$ are the uncertainties on the Cartesian coordinates
of the dipole.

\subsection {Dipole Calculation. Filling the Zone of Avoidance}  

The region of the sky close to the galactic plane, roughly bounded by
$|b|<20^\circ$, is not sampled by our cluster set. The effect on the dipole 
calculations of the ZoA, which amounts to approximately 
a quarter of the sky, is estimated by relying on the Monte Carlo approach 
described in the preceding section, of generating a large number of synthetic 
cluster sets. For a given observed set of $N_c$ clusters, we produce $N_{sets}$ 
of $1.25N_c$ clusters, where the additional $0.25N_c$ ``cloned'' clusters are 
assigned random coordinates in the ZoA and distances $c z_{tf}$ which are random 
deviates of the distribution for the $N_c$ observed clusters. The assignment
of a peculiar velocity to the cloned clusters is approached in two
independent ways, providing outer boundaries to the estimate of the effect
of ZoA on the uncertainty of the dipole solution:

\noindent (i) In the first approach, a peculiar velocity is extracted from a 
Gaussian distribution of zero mean {\it with respect to the Local Group}, and 
dispersion $\sigma_{1d}=300$ \kms. This approach reduces the amplitude of any 
dipole signal that may be present in the $N_c$ cluster sample.
Assigning a peculiar velocity from a Gaussian distribution of zero mean
in the CMB reference frame would be more appropriate, on the basis of the 
results discussed in Section 6, but it would reinforce the match of the reflex 
cluster dipole with {\bf V}$_{cmb}$. The chosen approach, while not producing
reliable dipole parameters, will yield an upper limit to the uncertainty of 
the dipole determination arising from the ZoA undersampling bias.

\noindent (ii) In our second approach to estimating the peculiar velocity
of a random cluster in the ZoA, we resort to an independent data set of 
peculiar velocities: that provided by our sample of field late spiral
galaxies (SFI; see e.g. Giovanelli \etal 1994 for a description of the
sample). SFI  is slightly less deep than the SCI cluster sample,
but it contains a sufficient number of galaxies at $cz$ near 9,000 \kms
for the purposes of this exercise.
It includes 680 galaxies with galactic latitudes lower than
$30^\circ$. Once a ZoA cluster is randomly assigned position $\hat {\bf r}_i$
and distance $c z_{tf,i}$, SFI galaxies within 2000 \kms of the cluster
are selected and ranked by the distance between galaxy $j$ and the cluster,
$d_j=\sqrt{(c z_{tf,i}{\bf \hat r}_i - c z_{tf,j}{\bf \hat r}_j)^2}$ . The cluster
is then assigned a peculiar velocity $v=\sum_j v_j d_j^{-1}/\sum d_j^{-1}$,
where the sum is over the nearest 15 galaxies to cluster $i$.
This approach preserves the characteristics of the large--scale peculiar
velocity field and yields the best possible estimate of the cluster dipole, 
corrected for the effect of ZoA undersampling.

\subsection {Dipole Calculation. Results}  



Table 2 displays values of the apex coordinates of dipole solutions
of the reflex motion of the LG with respect to the cluster set, calculated 
for a variety of subsets and processing options. 

Solution sets are computed for three main subsamples: (a) all clusters
together (sets 1 through 3); (b) clusters farther than $cz_{tf}=3000$ \kms
(sets 4 through 6); (c) clusters within $cz_{tf}=6000$ \kms (sets 7 
through 9). Solutions are computed for {\bf in} cluster galaxy samples;
the analogous ones for the {\bf in+} samples do not differ in a significant
way from those tabulated.
Computations are carried out with the following different approaches:
without correcting for the undersampling in the ZoA (sets 1, 4,  and 7); 
correcting for the ZoA undersampling by simulating mock
clusters with zero mean peculiar velocity with respect to the LG
(sets 2, 5 and 8); correcting for the ZoA undersampling
by simulating mock clusters of peculiar velocities derived from those
of the field (SFI) galaxies in the ZoA neighborhood (sets 3,6 and 9).
Furthermore, each solution is computed separately weighing each 
cluster by setting $s_i\equiv 1$ (columns 3 and 4, labeled 'Case $a$') and 
by setting $s_i\epsilon_i^2\equiv 1$ (columns 5 and 6, labeled 'Case $b$'). 
Tabulation of the dipole solutions includes the description of the
solution set (col. 1), the number of clusters per solution set (col. 2),
the modulus of the dipole vector ${\bf V}_d$ and its apex galactic 
coordinates (cols. 3, 4 and 5), with estimates of errors. The error
estimates were carried out by producing $N_{sets}=1000$ synthetic
versions of each sample, as discussed in Section 6.2. The solutions 
we consider the most robust are numbers 3, 6 and 9 

Figure 4 displays the apices of some of the solution sets in Table 2,
namely those for sets 3, 6 and 9. In each case we display both the $a$
(left--hand column) and the $b$ solution (right--hand column). On a grid 
of galactic coordinates centered at $(l,b)=(180,0)$, with $l$ increasing 
right to  left, we plot the dipole apices of each of the $N_{sets}=1000$ 
synthetic cluster sets corresponding to each sample. Contours at the 63\% (one 
sigma) and 95\% (2 sigma) confidence levels are plotted. The dipole solution 
in each case (the center of those contours) is also entered at the top right
of each panel, in the following order: dipole velocity, apex longitude and 
latitude. The apex of the motion of the LG with respect to the CMB is plotted 
as a large filled circle, and the apex of the LG motion in the Lauer \&
Postman solution is plotted as a large, crossed circle.

Figure 5 displays the apices of solution sets 4, 5 and 6, all for the
same subsample of clusters with $cz_{tf}>3000$ \kms. This figure illustrates
the effect of different approaches to correcting for the undersampling in the
ZoA.

We note immediately that the dipole of the LG motion with respect to the 
various sets of clusters is in substantial agreement with the CMB solution, 
while the Lauer \& Postman solution is excluded with a high level of confidence. 
The inclusion of clusters in the ZoA with random peculiar velocities 
averaging zero, as shown in panels (c) and (d) of Figure 5, increases the 
uncertainty of the dipole solution and raises its apex to higher positive 
latitudes. On the other hand, the adoption of synthetic clusters in the ZoA 
with peculiar velocities as inferred from those of neighboring field galaxies 
suggests a tightening of the quality of the solution, an indication that both 
the set of clusters of galaxies and that of field galaxies reflect the same 
peculiar velocity field. 

As indicated in Table 1, nearby clusters exhibit relatively large peculiar 
velocities with respect to the CMB, sharing that characteristic with the LG. 
Their exclusion, as shown in panels (c) and (d) of Figure 4, thus yields 
dipole solutions that approach even more closely the CMB solution. 
While the uncertainties are larger for solutions 4--6, they suggest that
{\it the dipole moment of the LG motion
with respect to clusters at $c z_{tf} > 3000$ \kms matches well
the CMB dipole; this result is consistent with the assumption that, as
a whole and apart from solutions which are purely geocentric, that set 
of clusters is at rest with respect to the CMB.}. This result is in good
agreement with an analogous study carried out using field spirals (Giovanelli
\etal 1998), which yields strong indication for the fact that at distances
on the order of  5--6000 \kms, convergence appears to have been achieved.
Given the magnitude of the uncertainties displayed in Table 2, it is however
fair to say that the cluster data set alone leaves room for 
the possibility that up to one--third of ${\bf V}_{cmb}$ arising outside 
the volume subtended by the cluster sample. The field galaxy sample does
however pose tighter constraints, and we are inclined to believe that
the combined evidence of field and cluster dipoles establishes a good
case for convergence within 9000 \kms.

In Figure 6, we plot all the dipole solutions listed in Table 2 
in two of three possible stereographic projections, i.e. in $(l,b)$ and 
$(|{\bf V}_d|,b)$, approximated as Cartesian coordinates. The CMB dipole is 
plotted as a filled circle and the Lauer \& Postman solution of the LG dipole 
as a crossed circle.

\subsection {Bulk Flows}  

The difference $({\bf V}_{cmb} - {\bf V}_d)$ for any of the solutions in
Table 2 yields the bulk flow motion of the corresponding sample with respect 
to the CMB. Because many of the cluster dipole solutions match 
so closely the CMB dipole, resulting bulk flows are quite modest, and
their directions largely unconstrained. The most robust estimates of
the bulk flow for the total cluster sample is
310$\pm120$ \kms, towards $(337^\circ,-15^\circ)\pm23^\circ$ (Case $a$) and
151$\pm120$ \kms, towards $(295^\circ,+28^\circ)\pm45^\circ$ (Case $b$), values
which are obtained for solution 3. For solution 6 of the $cz_{tf}> 3000$ \kms 
subsample, the bulk flow is not significant for case `a' and a barely marginal
$165\pm150$ towards $(278^\circ,-7^\circ)\pm45^\circ$ for case `b'.
 For solution 9 of the $cz_{tf}< 6000$ \kms subsample, the two bulk flow solutions 
are respectively 336$\pm144$ \kms, towards $(348^\circ,-20^\circ)\pm25^\circ$
and a marginal 131$\pm90$ \kms, towards $(325^\circ,+62^\circ)\pm60^\circ$
At the 95\% confidence level or better, these data exclude the existence of 
a bulk flow of amplitude 350 \kms or larger, centered on the LG, involving the 
equal volume--weighted contents of a sphere of 6000  \kms radius, or of 
the contents of a spherical shell of radius between 3000 and --9000 \kms.

\section {Summary}                                     

We have analyzed the peculiar velocity field as described by a set of
24 clusters and groups of galaxies at $cz$ between 1000 and 9200 \kms.
The following results emerge from our study:

\noindent $\bullet$ The peculiar velocity field as outlined by these
objects is rather quiescent. No velocities in excess of 600 \kms, with
respect to the CMB reference frame, are observed. This is in marked
contrast with previous results.

\noindent $\bullet$ The main supercluster structures within the reach
of the sample, such a the Coma supercluster and the Pisces--Perseus
supercluster, exhibit global deviations from Hubble flow in the CMB
reference frame that cannot
be distinguished from null, with uncertainties of less than 150 \kms.
Global flows of those structures at velocities in excess of 400 \kms
can be excluded at the 99\% confidence level.

\noindent $\bullet$ The dispersion of the line--of--sight peculiar velocity 
distribution function of clusters, as measured in the CMB reference frame,
is $\sigma_{1d}=300\pm 80$ \kms.

\noindent $\bullet$ We do not find evidence of a `Hubble bubble', 
i.e. a geocentric deviation from Hubble flow with amplitude of 6.6\% 
within $cz\sim 7000$ \kms, as reported by Zehavi \etal (1998). Our sample
does however barely straddle the edge of the bubble and has low statistical
significance. We can thus neither corroborate nor refute the Zehavi \etal
result.

\noindent $\bullet$ The dipole of the reflex motion of the LG with respect 
to the cluster set approaches closely the vector of the CMB dipole.
When the dipole is computed with respect to the clusters that are farther
than $cz\sim 3000$ \kms, the two coincide within the errors. This result
suggests that the convergence depth of the local universe is largely approached
within the limits of the cluster sample, i.e. the more distant clusters
in our sample populate a shell globally at rest with respect to the CMB.
The cluster set alone is however insufficient to exclude that up to 
1/3 of ${\bf V}_{cmb}$ may arise outside the volume subtended by the
sample. A parallel study which uses field spirals is in agreement
with the cluster data and reinforces the indication of convergence 
within 6--9000 \kms.

\noindent $\bullet$ The Lauer \& Postman dipole solution is excluded as 
possible by our data at better than the 99\% confidence level. The bulk 
flow of the contents of a sphere of radius 6000 \kms, centered on the LG, 
is small. Its amplitude is less than 300 \kms and its poorly constrained 
apex is in the general direction of $l=320^\circ$.

\acknowledgements

The results presented in this paper are based on observations carried out at
the Arecibo Observatory, which is part of the National Astronomy and 
Ionosphere Center (NAIC), at Green Bank, which is part of the National Radio 
Astronomy Observatory (NRAO), at the Kitt Peak National Observatory (KPNO), the 
Cerro Tololo Interamerican Observatory (CTIO), the Palomar Observatory (PO), 
the Observatory of Paris at Nan\c cay and the Michigan--Dartmouth--MIT 
Observatory (MDM). NAIC is operated by Cornell University, NRAO  by  
Associated Universities, Inc., KPNO and CTIO by Associated Universities 
for Research in Astronomy, all under cooperative agreements with the National 
Science Foundation. The MDM Observatory is jointly operated by the University 
of Michigan, Dartmouth College and the Massachusetts Institute of Technology 
on Kitt Peak mountain, Arizona. The Hale telescope at the PO is operated by 
the California Institute of Technology under a cooperative agreement with 
Cornell University and the Jet Propulsion Laboratory.  
This research was supported by NSF grants AST94--20505 and AST96--17069 to RG, 
AST95-28860 to MH and AST93--47714 to GW.

\newpage

\begin{figure}
\plotfiddle{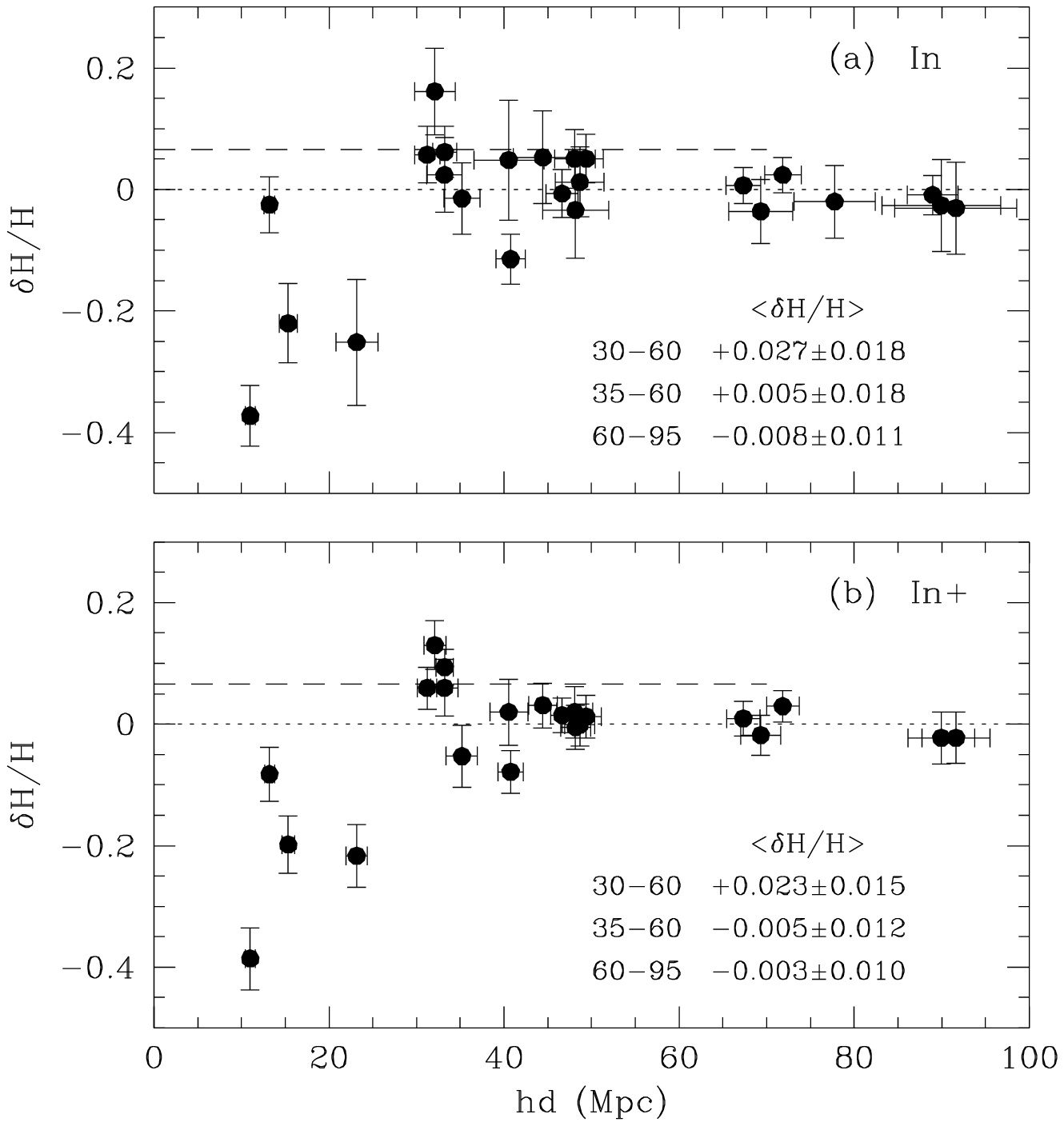}{5in}{0}{90}{90}{-255}{0}
\caption{Deviations from Hubble flow plotted versus TF distance
for the clusters listed in Table 1, separately for the {\bf in} (panel
$a$) and for the {\bf in+} (panel $b$) samples. The horizontal dashed
lines identify the acceleration of 6.6\% in the Hubble flow within
$hd=70$ Mpc claimed by Zehavi \etal (1998). Average values of 
$\delta H/H$ are inset for three different windows in $hd$.}
\end{figure}

\begin{figure}
\plotone{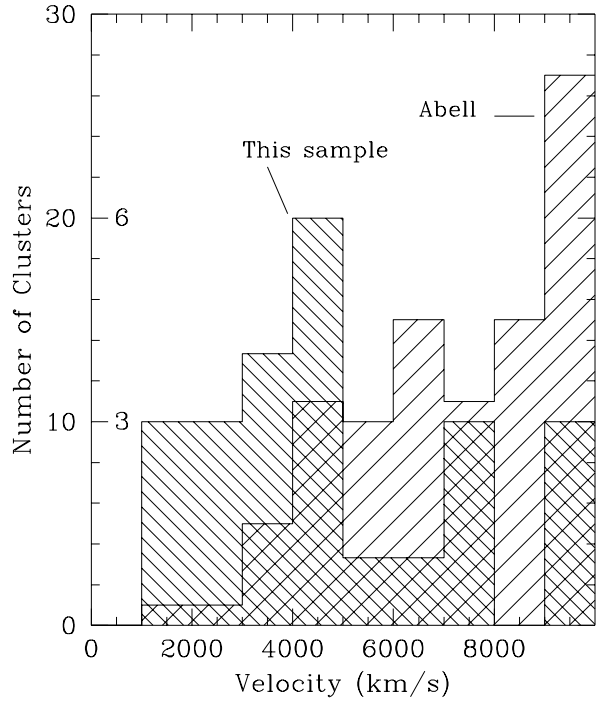}
\caption{Histograms of the number of clusters in our sample {\it vis--a--vis}
that in the Abell cluster catalog. The outer vertical scale applies to the 
latter, the inner one to the former.}
\end{figure}

\begin{figure}
\plotone{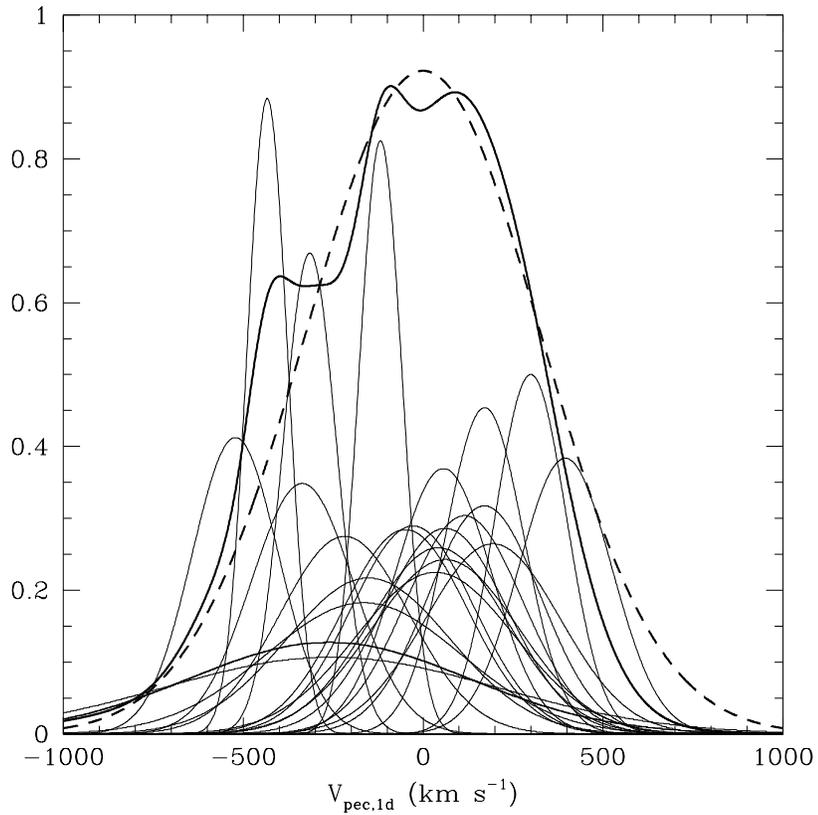}
\caption{Line--of sight peculiar velocities, measured in the CMB reference
frame, for the {\bf in+} samples 
of each cluster listed in Table 1, plotted as equal area Gaussians of 
dispersion equal to the uncertainty on each measured peculiar velocity.
The heavy--trace line is a scaled sum of the individual Gaussians.
The dashed line is a Gaussian of dispersion 325 \kms.}
p\end{figure}
\begin{figure}
\plotone{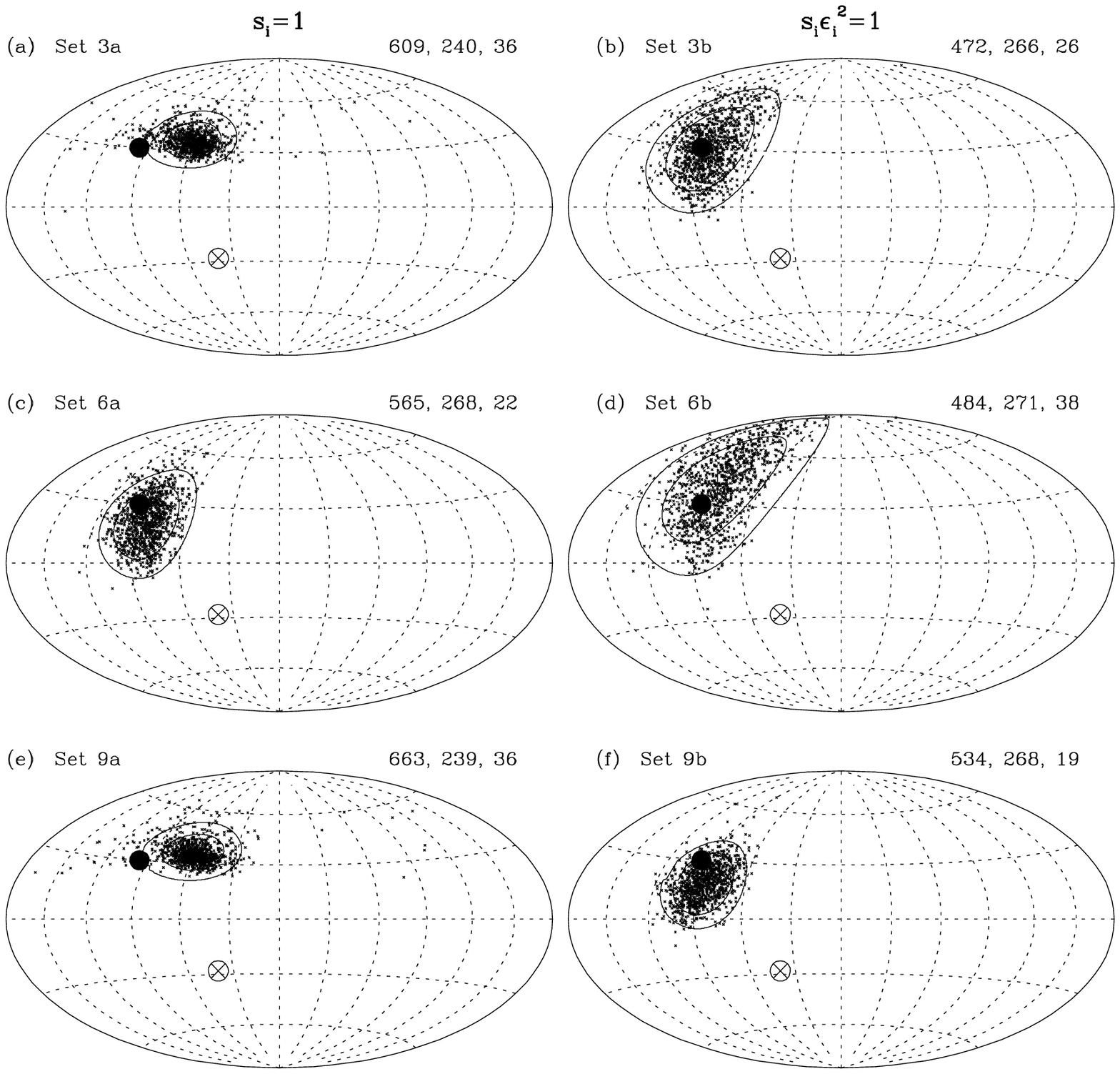}
\caption{Dipole solutions of $N_{sets}=1000$ synthetic data sets
with the characteristics of solutions 3, 6 and 9 of Table 2. The
coordinates of the Aitoff projections are galactic, centered at
$(l,b)=(180^\circ,0^\circ)$. The left--hand side panels correspond
to solutions estimated with weights as in approach (i)
in Section 7.2), while in right--hand panels weights are as in
approach (ii) in Section 7.2. The number of the solution on top left
of each panel refers to the line number in Table 2. The three numbers
on top right of each panel are the amplitude, longitude and latitude
of the dipole apex. The large, filled circle in each plot identifies
the apex of the CMB dipole, and the apex of the LG motion in the Lauer 
\& Postman solution is plotted as a large, crossed circle.}
\end{figure}

\begin{figure}
\plotone{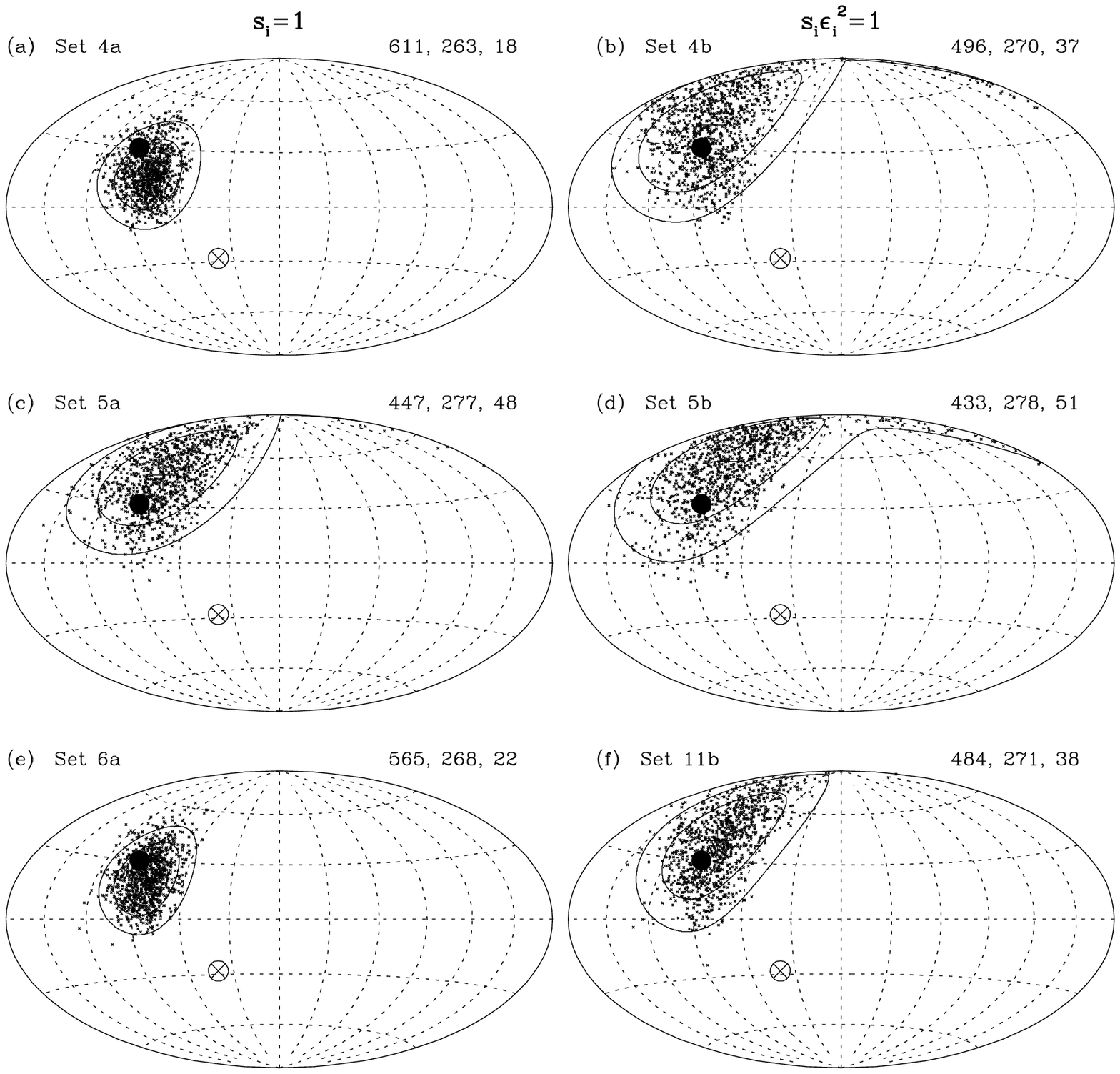}
\caption{Dipole solutions of $N_{sets}=1000$ synthetic data sets
with the characteristics of solutions 4, 5 and 6 of Table 2. The
coordinates of the Aitoff projections are galactic, centered at
$(l,b)=(180^\circ,0^\circ)$. The left--hand side panels correspond
to solutions estimated with weights as in approach (i)
in Section 7.2), while in right--hand panels weights are as in
approach (ii) in Section 7.2. The number of the solution on top left
of each panel refers to the line number in Table 2. The three numbers
on top right of each panel are the amplitude, longitude and latitude
of the dipole apex. The large, filled circle in each plot identifies
the apex of the CMB dipole, and the apex of the LG motion in the Lauer
\& Postman solution is plotted as a large, crossed circle.}
\end{figure}

\begin{figure}
\plotone{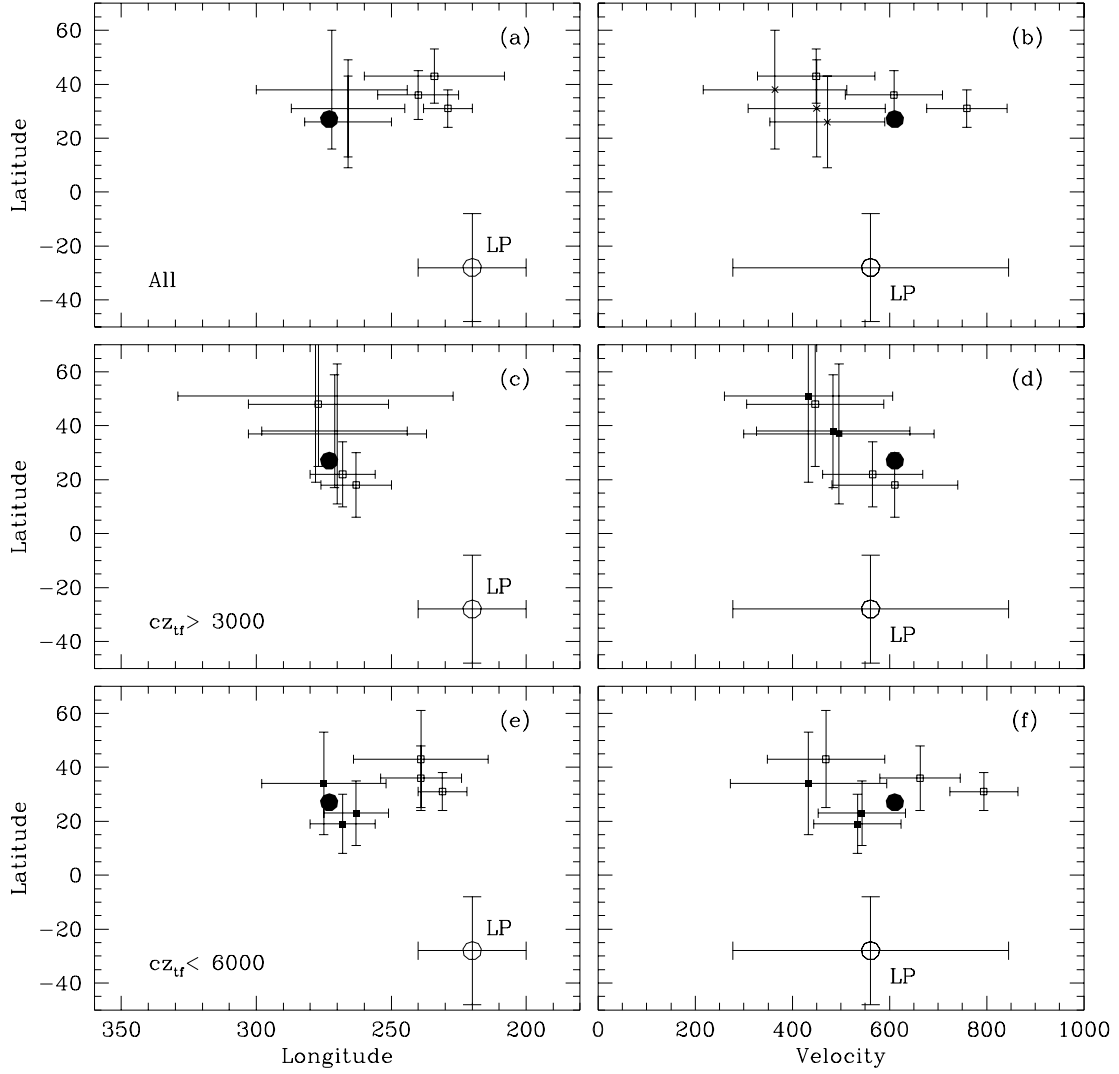}
\caption{Dipole solutions listed in Table 2 are shown as plain error bars,
in pairs of stereographic displays in $(l,b,|{\bf V}_d|)$,
respectively for sets 1--3 (panels a and b), 4--6 (panels c and d) and
sets 7--9 (panels e and f). Solutions of type `a', estimated by setting 
$s_i\equiv 1$, are plotted as small unfilled squares, while those of
type `b', estimated by setting $s_i\epsilon_v^2\equiv 1$, are plotted as 
filled squares. The large filled circle is the CMB dipole
and the crossed circle is the Lauer \& Postman dipole solution. }
\end{figure}

\vfill
\end{document}